\documentclass[aps,prl,twocolumn,amsfonts,showpacs]{revtex4} 
\usepackage{epsfig,amsopn}
\usepackage{graphicx}

\begin{document}
\def\be{\begin{equation}}
\def\ee{\end{equation}}
\def\bea{\begin{eqnarray}}
\def\eea{\end{eqnarray}}
\def\bml{\begin{mathletters}}
\def\eml{\end{mathletters}}
\def\l{\label}
\def\av#1{{\langle  #1 \rangle}}
\def\dav#1{{\overline{\langle  #1 \rangle}}}
\def\tnd{\rightarrow}

\title{Dynamics of a Disordered, Driven Zero-Range Process in One
Dimension}
\author{Kavita Jain and Mustansir Barma}
\affiliation{Department of Theoretical Physics, Tata Institute of
Fundamental Research, Homi Bhabha Road, Mumbai 400 005, India.}
\begin{abstract}
We study a disordered, driven zero range process which models a closed
system of attractive particles that hop with site-dependent rates
and whose steady state shows a condensation transition with
increasing density. We characterise the dynamical properties of the
mass fluctuations in the steady state in one dimension both analytically and 
numerically and show that there is a dynamic phase transition in
the density-disorder plane. We also
determine the form of the scaling function which describes the growth
of the condensate as a function of time, starting from a uniform
density distribution. 
\end{abstract}

\maketitle

Since the state of a system with an ordered phase in the infinite time 
limit is typically very different from that far from the steady state, 
different processes and time scales may govern the dynamics in the steady 
state and the relaxation towards it. Quenched disorder can strongly 
affect the dynamical properties in both situations; in particular, it may 
give rise to new dynamic universality classes. 
In the absence of a general framework for analyzing nonequilibrium, 
disordered systems, it is evidently of interest to develop a detailed 
understanding of these changes using simple models.

We address the above issues in a disordered, driven zero range process
(ZRP) which is a stochastic lattice model of interacting particles.
In this process, a site can be occupied by an arbitrary number of
unit-mass particles. Interparticle interactions are 
modeled by allowing the 
hop-out rate of a particle to depend on the mass at the site it leaves; in
general, these rates may even be site dependent. Remarkably, for any
choice of rates, the steady state of this model can be found exactly
\cite{spitzer}. There has been a surge in interest in the ZRP
following the finding that 
this model can show a condensation transition in which
at high densities, a finite fraction of particles condense onto a
single site. This transition occurs in the steady state of the
conserved mass model for a wide choice of hopping rates \cite{evansR}.  
Recent work on the ZRP has been devoted to studying dynamical properties 
\cite{kf,dynm,prmn} and using it to 
develop a general understanding of nonequilibrium steady states
\cite{ness1,ness2}, besides modeling various physical systems
\cite{evansR,sand}. 

Here we consider a disordered, driven ZRP in one dimension in which
a particle hops forward at a rate which is independent of the
mass. Quenched disorder is modeled by choosing site-dependent hopping rates
drawn from a distribution. The steady state of such a ZRP has been 
shown to exhibit a phase transition, from a 
low density, homogeneous phase to a high density phase with
a condensate at the site with the lowest hopping rate \cite{kf,bec}.
Interestingly, by regarding sites as particles and masses as
hole clusters, this model maps exactly onto a simple traffic model of
cars (particles) with different preferred speeds on a single-lane
highway, with no possibility of overtaking. At low densities, an
infinite headway appears in front of the slowest car, corresponding to
the condensate in the ZRP \cite{kf,bec}.  

Our results pertain both to the dynamical properties in the steady
state and to the manner in which the system relaxes to it. While the
former concerns the motion and the decay of the density fluctuations
about the mean in the bulk of the system, the latter involves the
transfer of a macroscopic amount of mass from the bulk of the
system to the globally slowest site. We find that different time scales
govern these two processes. Our main results are summarized below. \\
{\it{Steady state dynamics:}} We calculate the speed of density
fluctuations in the steady state and identify the regimes in the
density-disorder plane in which this speed vanishes, signalling
a dynamic phase transition. Our Monte Carlo simulations show that when this
speed is nonzero, the dynamic behavior remains the same as that of a pure
system, while it changes if the speed vanishes. \\
{\it{Relaxation to the steady state:}} We give an analytical argument
for the form of the scaling function which describes the temporal
growth of the condensate
starting from a uniform density distribution, and present numerical
evidence to support our results. The dynamic exponent is deduced from
the growth law via a scaling argument and agrees with earlier results
based on a deterministic traffic model \cite{kf,newell,krugR}. 


The ZRP involves $M$ particles on a ring of size
$L$ with an arbitrary number of particles allowed at any site. A particle
hops out of a randomly selected site $k$ to site $k+1$ with 
quenched rate $w_{j}(k)$ where the subscript $j$ is the index which
ranks the rates in ascending order, with $j=1$ labeling the lowest rate.
The rate $w(k)$ is chosen to be independent of the mass at site $k$ so
that the system has on-site attractive interactions \cite{attr}.
These site-dependent hopping rates are chosen independently from a
common distribution  
\be
f(w)=\left[ (n+1)/(1-c)^{n+1} \right] \; \left(w-c \right)^{n} \;,\;w \in
[c,1] \;,
\ee
with $c, n > 0$. For this process, the probability of a configuration
$C \equiv \{m(1),...,m(L) \}$ in the steady
state is given by \cite{kf,bec} 
\be
P(C)= \frac{1}{Z} \prod_{k=1}^{L} \left( \frac{v}{w(k)}
\right)^{m(k)} \;, \l{pm}
\ee
with the constraint $\sum_{k} m(k)=M$. Here $Z$ is the partition
function, $m(k)$ is the mass 
at site $k$ and $v$ is the fugacity. The preceding equation gives 
the average mass $\mu_{j}=v/(w_{j}-v)$ at the site with hopping rate
$w_{j}$. Since the total number of particles is conserved, we have
\be
\rho=\frac{1}{L} \; \frac{v}{w_{1}-v}+\int_{c}^{1} dw \;
\frac{v}{w-v} \;\;f(w) \l{cons} \;, 
\ee
where $\rho=M/L$ is the density. The above equation implies that in
the thermodynamic limit, there exists
a finite critical density $\rho_{c}$ below which the fugacity
increases with density and above which $v$ gets pinned to the lowest 
hopping rate $c$. Thus, there is a phase transition from the low
density phase with mass of order unity at each site to a high
density phase with infinite mass at the site with the lowest hopping
rate \cite{kf,bec}. The critical point, given by 
$\rho_{c}=c\;(n+1)/n(1-c)$, is obtained   
from Eq.(\ref{cons}) on setting $v$ equal to $c$ in the integral. This
transition is analogous to Bose-Einstein condensation in the ideal
Bose gas or in a system of noninteracting bosons in a random
repulsive potential \cite{rbec}.

We begin with a discussion of the steady state dynamics, in
particular, the study of the statistical fluctuations of the
density about its average. In a steady state carrying a uniform
current $J$, these fluctuations are carried as kinematic waves whose
speed $v_{kin}$ is known to be given by $v_{kin}=\partial J /\partial 
\rho$ from a general hydrodynamic argument \cite{lw}. This speed plays an
important role in determining whether the quenched disorder
changes the dynamic universality class \cite{sitewise}. 
If $v_{kin}$ is nonzero, each density fluctuation encounters a
particular patch of disorder essentially only once in an infinite
system, as the probability of returning is exponentially small. Thus, 
the noise arising from the different patches of disorder is
essentially uncorrelated in time, and we would then expect
the kinematic wave to decay with the Kardar-Parisi-Zhang (KPZ)
exponent \cite{kpz}. However, if $v_{kin}$ vanishes, then
this argument fails and we would expect disorder to change the dynamic
universality class.

In order to determine the kinematic wave speed, we use $J$=$v$ and
Eq.(\ref{cons}) to obtain
\be
v_{kin}^{-1}= \frac{1}{L} \frac{w_{1}}{(w_{1}-v)^{2}}+
\frac{1}{L} \frac{w_{2}}{(w_{2}-v)^{2}}+ \int_{c}^{1} dw \; \frac{w
\;f(w)}{(w-v)^{2}} \l{kin} \;, 
\ee
where we have separated out the contributions from the two slowest
sites. The above expression for $v_{kin}^{-1}$ involves essentially
the sum of the 
mass variances $\sigma_{j}^{2}$ at sites with rate $w_{j}$ where 
$\sigma_{j}^{2}=v \;w_{j}/{{(w_{j}-v)^{2}}}$ is obtained 
for all $j$ and $\rho$ using Eq.(\ref{pm}). (However, this
expression for variance is invalid at the slowest site for $\rho 
> \rho_{c}$, as explained below.) For a given set $\{ w \}$ and fixed
density, we 
solved Eq.(\ref{cons}) numerically to determine the fugacity $v$ which
is used to find $\mu_{j}$ and $\sigma_{j}^{2}$. The $L$ dependence of
the mass and the variance was then found by averaging over a large
number $(\sim 10^{6})$ of disorder configurations.
Our findings for the mean mass $\mu_{j}$ and the fluctuations
$\sigma_{j}$ are summarized below. 

For $\rho < \rho_{c}$, each site supports mass of order
unity with fluctuations of the same order. 

For $\rho=\rho_{c}$, both $\mu_{j}$ and $\sigma_{j}$  
at a site with ordering index $j$ grow as 
$(L/j)^{1/(n+1)}$. This can be seen by using $v(\rho_{c}) \simeq c$
and changing $w$ to a uniformly distributed variable $u$ defined by
$w-c=(1-c) \; u^{1/(n+1)}$.  

For $\rho > \rho_{c}$, the behavior of $\mu_{j}$ and 
$\sigma_{j}$ in a typical disorder configuration remains the same as
that at the critical point for $j \geq 2$. However, the
disorder-averaged variance $\overline{\sigma_{j}^{2}}$ for 
$j=2$ grows with $L$ at the same rate as that for $j=1$, and is larger than
at any other site in the rest of the system, even though the second
slowest site does not support a condensate.
To understand this surprising feature, we numerically studied the 
distribution $P(x_{j} \equiv w_{j}-v,L)$ for various $j$ and $L$. For 
the slowest site, this distribution approaches a delta function centered about
$1/L$. Thus the expression for $\sigma_{j}^{2}$ quoted above
breaks down for $j=1$ since it predicts macroscopic fluctuations at 
this site, implying the invalidity of the grand 
canonical ensemble. To determine the variance at this site, we 
use the sum rule $\sigma_{1}^{2}=\sum_{j=2}^{L} \sigma_{j}^{2}$ which follows 
from mass conservation and the product measure form of the steady state.
Our numerical data shows that for $j > 1$ and large $L$, the
distribution $P(x_{j}, L)$ is of the form    
\be
P(x_{j}, L) \approx L^{1/{(n+1)}} \; X_{j}\left[(x_{j}-\epsilon_{j}) \;
L^{1/{(n+1)}} \right] \Theta(x_{j}-\epsilon_{j}) \;,\l{dist}
\ee
where $\Theta$ is the Heaviside step function, $\epsilon_{j}$ is of order 
$1/L$ and $X_{j}$ is the scaling function appearing in the
distribution ${\cal {P}}(w_{j}-w_{1},L)$ of  
variable $w_{j}-w_{1}$ in $L$ trials. This distribution can be
calculated and is of the same scaling form as Eq.(\ref{dist}) with
$\epsilon_{j}=0$; the scaling function $X_{j}(x)$ is found to grow as
$x^{j-2}$ for $x \ll 1$ and 
decays as $(1-x)^{n}$ as $x \tnd 1$. Note that $X_{j}(x)$ approaches a nonzero
value as $x \tnd 0$ for $j=2$.
Using this scaling function in Eq.(\ref{dist}), we find that 
$\overline{\sigma_{j}^{2}} \sim (L/j)^{2/(n+1)}$ for $j >2$, while for $j=2$
it is of the order $L^{(n+2)/(n+1)}$. Since the contribution of
$\overline{\sigma_{2}^{2}}$ dominates the rest in the sum rule, we find 
\be
\overline{\sigma_{1}^{2}} \sim \overline{\sigma_{2}^{2}} \sim
L^{(n+2)/(n+1)} \;.\l{fluct} 
\ee
The anomalously large value of $\overline{\sigma_{2}^{2}}$ is a consequence of
the nonzero probability for a near-vanishing difference between
the two lowest rates. 

We now return to Eq.(\ref{kin}) and determine $v_{kin}$ using the
results obtained above. 

For $\rho < \rho_{c}$, the first two terms in Eq.(\ref{kin}) 
are negligible in the thermodynamic limit,
while the integral is of order unity so that $v_{kin}$ is nonzero. Because of
the argument above, we expect that the kinematic waves decay with
the KPZ exponent in this phase. 

For $\rho=\rho_{c}$, there is a transition in the behavior
of $\partial \rho/ \partial v$ as $n$ crosses one
\cite{kf}. Integrating $\sigma_{j}^{2}$ over $j$, we find that  
\be
v_{kin}^{-1}(\rho_{c}) \sim \cases{ L^{(1-n)/(1+n)} & {, $n < 1$} \cr
                       (n+1)/(n-1)  & {, $n > 1 \;,$}
} \l{kincp}
\ee
which indicates a dynamic phase transition at $n=1$ in the
thermodynamic limit. For $n > 1$, the kinematic wave speed is
nonzero and we expect the universality class to be the same as for
$\rho < \rho_{c}$.
For $n \leq 1$, this speed is zero implying that the
transport of density fluctuations is anomalously slow. Thus, for $n \leq 1$,
disorder is expected to be relevant in changing the dynamical behavior 
from the KPZ universality class. 

For $\rho > \rho_{c}$, due to the condensate at the slowest
site, the first term in Eq.(\ref{kin}) diverges giving $v_{kin}=0$ for
all $n$. We consider the speed $v_{kin}^{\prime}$ in
the system excluding this site. In a typical disorder configuration,
$v_{kin}^{\prime}$ behaves as at the critical point and there is a
transition in the dynamical behavior at $n=1$. However, the 
disorder average of the inverse speed
$\overline{v_{kin}^{\prime^{-1}}}$ diverges for
all $n$, as can be seen using Eq.(\ref{fluct}) in Eq.(\ref{kin}).

\begin{figure}
\begin{center}
\includegraphics[scale=0.352,angle=270]{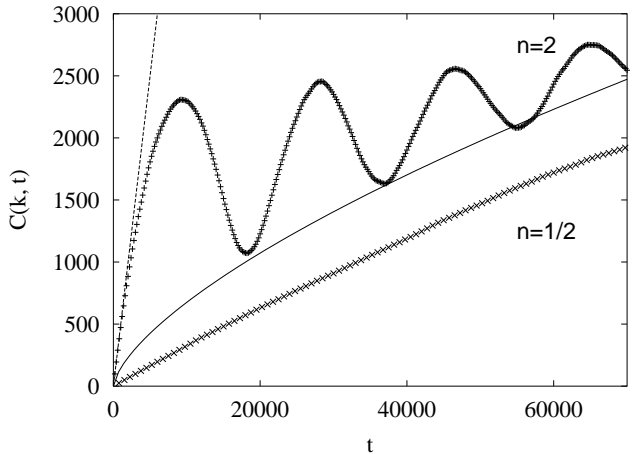}
\caption{Plot of the tagged particle correlation $C(k,t)$
vs. $t$ at the critical point showing the existence of kinematic waves
for $n > 1$. For $n=2$, the solid curve passing 
through the minima is a power law with exponent $2/3$ and the initial 
tangential straight line has a slope equal to $c=1/2$. For $n=1/2$,
the y-axis has been scaled down by a factor of $15$. The data
have been averaged over all $k$ for both values of $n$.} 
\label{cp}
\end{center}
\end{figure} 

We verified the above predictions by 
monitoring $C(k,t)=\av{H^{2}(k,t)}-{\av{H(k,t)}}^{2}$
where $H(k,t)$ is the number of particles that hop past site $k$
in a time interval $t$. 
In the traffic model, $C(k,t)$ gives the 
tagged particle correlation which measures the mean squared
displacement of a tagged particle $k$ around its mean position at
time $t$. In an infinite system, $C(k,t)$ increases linearly
with time and the slope gives the tagged diffusion constant. In a
finite system with periodic boundary conditions, if 
kinematic waves are present, $C(k,t)$ oscillates with time period 
$L/v_{kin}$ and its values at the minima are a measure of  
the decay of the wave. 

We measured $C(k,t)$ using Monte Carlo simulations in both the phases 
and at the critical point. Except at very short times,
$C(k,t)$ is found to be independent of $k$, as
explained below. In Fig. \ref{cp}, we show $C(k,t)$
as a function of time at the critical point for two values of $n$. For $n=2$,
it is found to oscillate,
with the values at minima growing as a power law in time with exponent
$2 \beta_{KPZ}=2/3$; at short times, it
increases linearly with slope equal to $c$. For $n=1/2$, there are no
oscillations and $C(k,t)$ continues to increase linearly with
slope equal to $c$. The tagged diffusion constant is the same for all $k$ and
is equal to that of the slowest particle for $\rho \geq \rho_{c}$, due to the
no-overtaking constraint in the traffic model. 
Since the slowest particle behaves as a free, biased random walker due to
the infinite headway in front of it, its diffusion constant equals $c$
for all $n$.


A different sort of kinetics governs the approach to the steady state
in the condensate phase in which, starting from a uniform density
distribution, the mass profile develops a singularity at the site with
the globally minimum hopping rate. In the initial stage, the particles
hop out of relatively 
fast sites quickly and get trapped temporarily at locally slow sites.
At moderately large times, one finds a finite density of large
aggregates at these slow sites, which relax by releasing their excess
mass to yet slower sites on their right. Thus, the
masses at slow sites first grow and then decay to their respective
steady state values, except at the slowest site where the mass monotonically
increases and then saturates \cite{prmn}.

For an analytical description of the above growth mechanism, it is
useful to consider a sequence of slowly relaxing sites on the right of
the slowest site. By a sequentially increasing label $\ell=1,2,...$,
we mark the set of 
sites which satisfy $\tau_{\ell} > \tau_{\ell-1}$ where $\tau_{\ell}$
is the relaxation time of $\ell$th such site. Denote its
position by $R_{\ell}$; it is evident that $R_{\ell}$ is a 
random variable which grows rapidly with $\ell$. Now $\tau_{\ell}$ can be
estimated by observing the following: (i) the mass at site $\ell$ grows as
$t^{\beta}$ till time scales of order $\tau_{\ell-1}$,
accumulating mass $\Delta m(R_{\ell}) \simeq \Delta \rho \; R_{\ell}$ where
$\Delta \rho \equiv (\rho-\rho_{c})$, and (ii)
when the excess mass has reached the site $\ell$, the region to its
left has relaxed to the true steady state. At this point, the mass at site
$\ell$ begins to decrease since the out current $J_{out} \approx
w(R_{\ell})$ exceeds the in current $J_{in}=c$ leading to $\tau_{\ell} \approx
\Delta m({R_{\ell}})/(w(R_{\ell})-c)$. 

The growth of mass at site $\ell$ can be described by
considering the distribution of mass $\Delta m(R_{\ell})$ at location
$R_{\ell}$ such that
$\tau_{\ell} > t$ and $\tau_{\ell^\prime} < t$ with $\ell^{\prime} < \ell$.
The probability of this event is $g((\Delta \rho \; R_{\ell})/t) 
\prod_{R_{\ell^{\prime}} < R_{\ell}} [1-g((\Delta \rho \;
R_{\ell^{\prime}})/t)]$ 
where $g(u)=u^{n+1}$ is the probability that $w-c  < u$. It follows
that the cumulative distribution $F(\Delta m, t)$ of
having mass up to $\Delta m$ at time $t$ has the scaling form 
\be 
F(\Delta m,t) \approx 1-\mbox{exp}(-b_{n} y^{n+2}) \;, \l{coll}
\ee
where $b_{n}$ is a constant and  $y=\Delta m/t^{\beta}$ is the scaling
variable with $\beta=(n+1)/(n+2)$. We have dropped the label
$\ell$ since we expect the same growth dynamics for all slow sites
including the slowest. Further, the average mass $\overline{\Delta
m_{1}(t,L)}$ at the slowest site at time
$t$ in a system of size $L$ is expected to follow the scaling form 
$\overline{\Delta m_{1}(t,L)} \approx
t^{\beta} H(t/L^{z})$ where $z=1/\beta$ is the
dynamic exponent \cite{prmn}. In the traffic model, starting from a 
homogeneous initial condition, the system approaches the steady state
via a coarsening process by which headway lengths grow . The above
expression for $\beta$ matches 
with the growth exponent for the typical headway length obtained using a 
deterministic model \cite{kf,newell,krugR}. 

\begin{figure}
\begin{center}
\includegraphics[scale=0.352,angle=270]{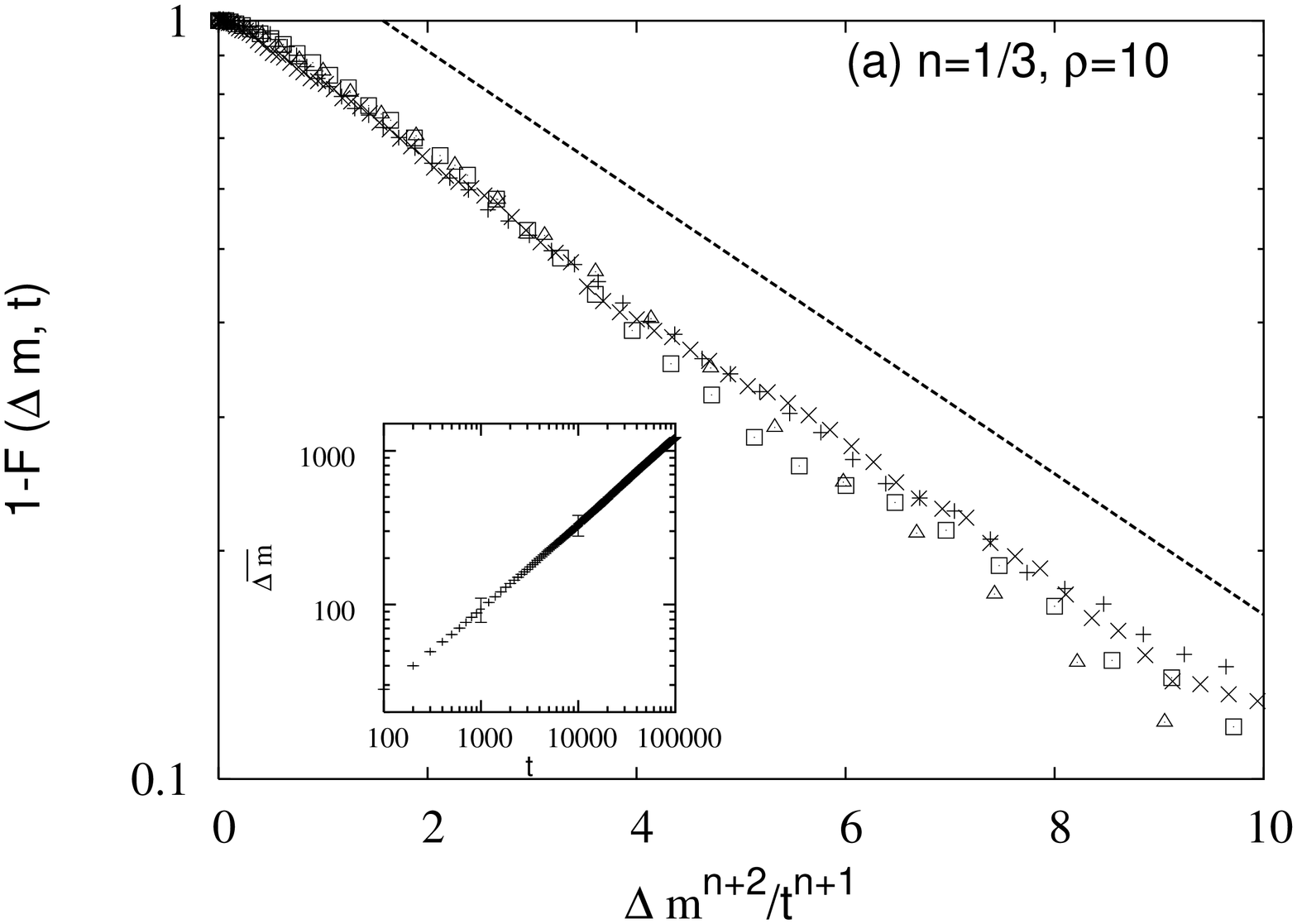}
\includegraphics[scale=0.352,angle=270]{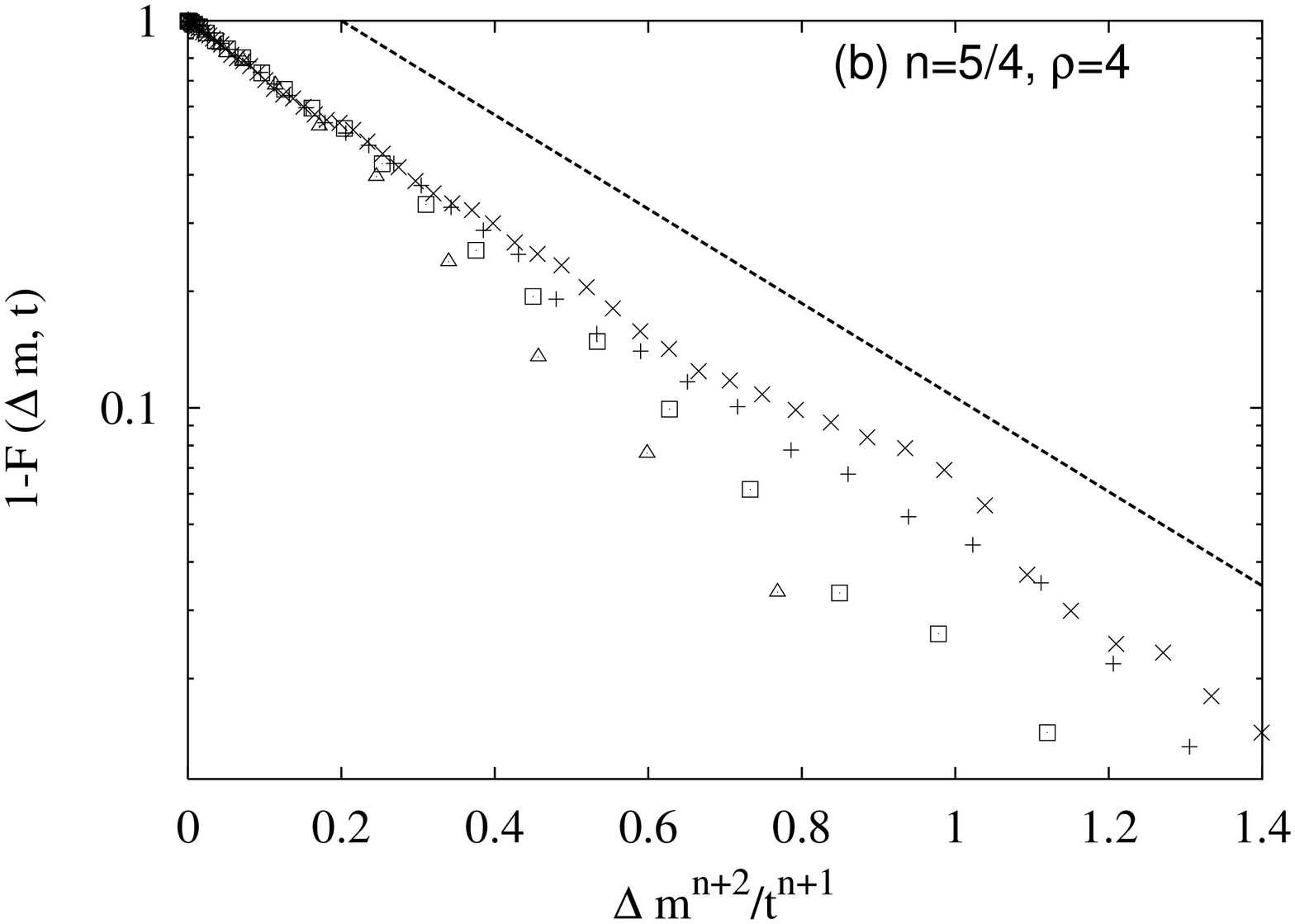}
\caption{Scaling collapse of the cumulative 
distribution in Eq.(\ref{coll}) for 
(a) $n$=$1/3$ (b) $n$=$5/4$ at $t$=$10^{4}(\triangle), 2 \times 10^{4}(\Box), 
4 \times 10^{4}(+)$ and $8 \times 10^{4}(\times)$ in the condensate phase. The 
data are obtained from the two slowest sites in $150$ disorder 
configurations, using $c=1/2$. The broken line is a guide to the eye. The inset
shows the temporal growth of average mass as a power law with exponent
$\beta$ given in the text.}
\label{coarsen}
\end{center}
\end{figure}

To test Eq.(\ref{coll}), we numerically measured the distribution
$F(\Delta m,t)$ 
at the slowest and the second slowest sites as a function of time in a
large system ($L \sim 5 \times 10^{4}$) for
various values of $n$. We used
slightly modified dynamical rules since it allowed us to access
longer times. We determined the out current $J(k,t)=w(k) \; s(k,t)$ at
site $k$ with occupancy probability $s(k,t)$ at time $t$ using the steady state
expression $s(k,t)=m(k,t)/(1+m(k,t))$, where $m(k,t)$ is the instantaneous 
mass at site $k$. The mass was updated using this modified expression for
current in the evolution equation of $m(k,t)$.
This is expected to be a good approximation at large times when the
system is close to steady state, and is exact in the steady state. We
also checked these simulation results against the original dynamics
for some cases and found that the results agree.
Figure \ref{coarsen} shows the collapse of
$\ln(1-F(\Delta m,t))$  
vs. $\Delta m^{n+2}/t^{n+1}$ onto a linearly decreasing curve, in 
accordance with Eq.(\ref{coll}), for two representative values of $n$.

To summarize, we studied a disordered, driven ZRP whose dynamical
properties in the steady state show a phase transition as a function
of disorder parameter $n$. The relaxation dynamics, by contrast,
depends smoothly on disorder parameter. The dynamic universality class in the
steady state was shown to remain same as for the pure system for $n > 1$,
while a complete characterisation of the new
universality class for $n < 1$ remains an interesting open question.

We thank D. Dhar for useful comments on the manuscript. K.J. acknowledges 
partial support from the Kanwal Rekhi scholarship administered by the TIFR 
Endowment Fund.

\begin{thebibliography}{10}
\bibitem{attr}
This choice of rates implies an attractive on-site interaction since
particle hopout rates are lower than those for noninteracting particles.
\end{thebibliography}
\end{document}